\documentstyle[prl,twocolumn,epsf,floats,aps]{revtex}

\newcommand{\beq}{\begin{equation}}
\newcommand{\eeq}{\end{equation}}
\newcommand{\bea}{\begin{eqnarray}}
\newcommand{\eea}{\end{eqnarray}}
\newcommand{\beano}{\begin{eqnarray*}}
\newcommand{\eeano}{\end{eqnarray*}}

\begin{document}
\draft

\twocolumn[\hsize\textwidth\columnwidth\hsize\csname
@twocolumnfalse\endcsname

\title{Thermal metal in network models of a disordered two-dimensional
superconductor}
\author{J. T. Chalker$^1$, N. Read$^2$,
V. Kagalovsky$^{3,4}$, B. Horovitz$^4$, Y. Avishai$^4$, and A. W. W.
Ludwig$^5$}
\address{$^1$Theoretical Physics, Oxford University, Oxford
OX1 3NP, United Kingdom\\ $^2$Department of Physics, Yale
University, P.O. Box 208120, New Haven, CT 06520-8120\\
$^3$Negev Academic College of Engineering, Beer-Sheva 84100, Israel\\
$^4$Department of Physics, Ben-Gurion University of the Negev,
Beer-Sheva 84105, Israel\\
$^5$Physics Department, University of California, Santa Barbara,
CA 93106-4030}
\date{\today}
\maketitle

\begin{abstract}
We study the universality class for localization which arises from
models of non-interacting quasiparticles in disordered
superconductors that have neither time-reversal nor spin-rotation
symmetries. Two-dimensional systems in this category, which is
known as class D, can display phases with three different types of
quasiparticle dynamics: metallic, localized, or with a quantized
(thermal) Hall conductance. Correspondingly, they can show a
variety of delocalization transitions. We illustrate this behavior
by investigating numerically the phase diagrams of network models
with the appropriate symmetry, and for the first time show the
appearance of the metallic phase.

\end{abstract}

 ]

The properties of quasiparticles in disordered superconductors
have been a subject of much recent interest. Within a mean
field treatment of pairing, the quasiparticles are noninteracting fermions,
governed by a quadratic Hamiltonian which may contain effects of
disorder in both the normal part and the superconducting gap
function. Such Hamiltonians are representatives of a
set of universality classes different from the three classes
which are familiar
both in normal disordered conductors and in the Wigner-Dyson random matrix ensembles.
A list of additional random matrix ensembles, determined by these new
symmetry classes, has been established relatively recently
\cite{az}. These additional random matrix ensembles describe
zero-dimensional problems, and are appropriate to model a small grain
of a superconductor in the ergodic
limit. In the corresponding higher-dimensional systems from the
same symmetry classes, there can be transitions between metallic,
localized, or quantized Hall phases for the quasiparticles
\cite{sfbn,khac,brad}. The associated changes in quasiparticle
dynamics must be probed by energy transport or (in singlet
superconductors) spin transport, rather than charge transport,
since quasiparticle charge density is not conserved \cite{sfbn}.
There are various possibilities for behavior, depending on the
particular symmetry class considered. These have been studied
theoretically using nonlinear sigma model methods \cite{sfbn},
numerically \cite{khac}, and in quasi-one dimensional models
\cite{bfgm}. An important question not addressed in such work so
far, and which will not be considered here, is whether the
self-consistent solution to the gap equation in the presence of
disorder affects the universal statistical properties of the
ensembles.

In this paper we present extensive numerical results on a
symmetry class with particularly rich phase diagram in two dimensions,
class D. The symmetry may be realized in superconductors with broken
time-reversal invariance, and either broken spin-rotation invariance
(as in d-wave superconductors with spin-orbit scattering) or
spinless or spin-polarized fermions (as in certain p-wave states).
The nonlinear sigma model for class D \cite{az}
has been shown, in the two-dimensional case, to flow under the
renormalization group to weaker values of the coupling constant
\cite{bundschuh,sf2,readgr,bsz}. The coupling constant is
proportional to the inverse of the thermal conductivity of the
superconductor, and this flow implies that there is a phase in
which there is a nonzero (indeed, diverging \cite{sf2}) density of
extended fermion eigenstates at zero excitation energy. A
superconductor described by this model would be in a thermal metal
phase. We will refer to such a phase simply as a metallic phase.
In addition, a phase with localized quasiparticles is a natural
possibility, and---since time-reversal symmetry is broken---so is
one with quantized Hall conductance. Our aim in the following is
to investigate the appearance of these phases in simple models.

\begin{figure}
\epsfxsize=2.375in \epsfxsize=3.25in
\centerline{\epsffile{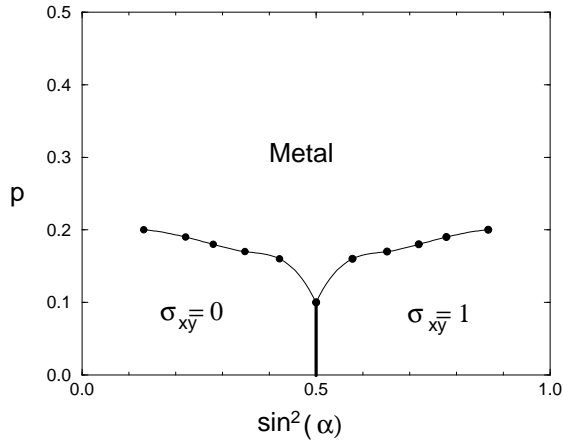}} \vspace{0.0in} \caption{The
phase diagram of the CF model obtained from our numerical calculations.} \label{phasediag}
\end{figure}

As our starting point, we take versions of the network model
\cite{cc} for a single-component fermion, which we specify in
detail after first summarizing our findings. Disorder appears in
the network model in the form of random scattering phases, and the
symmetries of class D restrict scattering phases to the values $0$
and $\pi$. Remarkably, within this framework, different particular
forms of disorder result in quite distinct physical behavior. We
discuss three alternative choices. The first of these (CF) was
introduced in work by Cho and Fisher \cite{cf} with the intention
of modeling the two-dimensional random bond Ising model (RBIM),
which possesses a fermion representation with the symmetries of
class D. In fact, as noted subsequently \cite{chothes,grl}, a
precise mapping of the Ising model leads to a second version of
the network model, which we label RBIM. In both these models,
scattering phases with the value $\pi$ appear in correlated pairs.
A third model (also discussed in Ref.\,\cite{bsz}), which we
denote by O(1), arises naturally if one instead takes scattering
phases to be independent random variables. Each model has two
parameters: a disorder concentration, $p$ ($0\leq p \leq 1$), and
a tunneling amplitude \cite{cc}, $\alpha$ ($0 \leq \alpha \leq \pi/2$),
which controls the value of the (thermal) Hall conductance at
short distances. Our phase diagram for the CF model in the
$(p,\alpha)$ plane is shown in Fig.\,\ref{phasediag}. It contains
a region of metallic phase,
and two distinct localized phases, which can be identified with
the ordered and disordered phases of the RBIM, or as regions with
different quantized Leduc-Righi (thermal Hall) conductivities. As
a consequence, three potentially different critical points occur:
an insulator-to-insulator quantum-Hall--type transition; an
insulator-to-metal transition; and a multicritical point at which
all three phases meet. This phase diagram has the form proposed
generically for class D in Ref.\ \cite{sf2}. In contrast, neither
the RBIM nor the O(1) model supports all three phases: arguments
that the metallic phase cannot appear in RBIMs with real Ising
couplings are given in Ref.\ \cite{rl}; while in the O(1) model we
find no localized phase, in striking distinction to all network
models studied previously. We show below how these differences can
be understood by solving the models in one dimension and by
considering them in two dimensions at weak tunneling.

\begin{figure}
\epsfxsize=2.375in \centerline{\epsffile{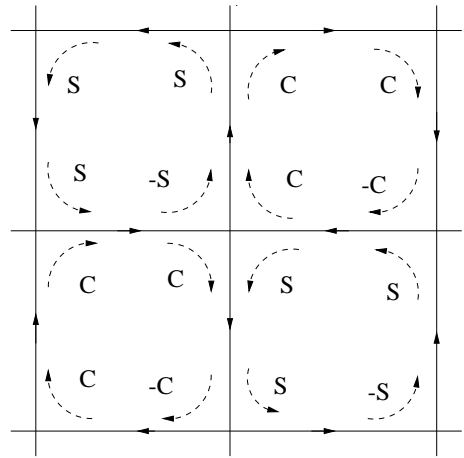}}
\vspace{0.1in} \caption{The network model. Values of the
scattering matrix elements, $\pm \cos(\alpha)$ and
$\pm\sin(\alpha)$, at nodes on each sublattice are indicated for
$p=0$ schematically by $\pm C$ and $\pm S$.} \label{network}
\end{figure}

All these models represent coherent
propagation of quantum-mechanical flux on a square lattice
of directed links which meet at nodes,
as illustrated in
Fig.\,\ref{network}.
Plaquettes of the lattice can be divided into two sets, according
to the direction of circulation around them. For general values of
$\alpha$, all plaquettes are coupled, but for $\alpha=0$ the
system separates into uncoupled plaquettes with clockwise
circulation, while for $\alpha=\pi/2$ it consists of uncoupled
anticlockwise plaquettes. Disorder is introduced via a phase
$\phi_l$ associated with each link, $l$. To make clear the
constraints imposed in class D, recall that a Bogoliubov-de Gennes
Hamiltonian with this symmetry may be written in terms of a purely
{\it imaginary} Hermitian matrix \cite{az}. The corresponding time
evolution operator is purely real, restricting the generalized
phase factors to be O($N$) matrices for a model in which
$N$-component fermions propagate on links, and to the values $\pm
1$ for $N=1$, the case we treat. It is useful to consider the
gauge-invariant total phase, modulo $2\pi$, accumulated on passing
around each elementary plaquette. In place of individual link
phases, randomness can be characterised by the positions of flux
lines which thread a subset of plaquettes, adding $\pi$ to their
phases.

The models we study and the important distinctions between them
are as follows. The CF model has transfer matrix
tunneling parameters chosen
negative with probability $p$, and positive with probability $1-p$
\cite{cf}. In consequence, flux lines appear in pairs at a 
node with probability $p$: both
members of a pair pass through plaquettes with the same
circulation, but different pairs may belong to plaquettes with
opposite circulation.
The RBIM similarly has a pair of flux lines introduced at each
node with probability $p$, but with the difference that all flux
lines thread plaquettes of the same circulation \cite{grl,rl}.
Finally, the O(1) model has link phase factors chosen negative
with probability $p$ and positive with probability $1-p$, so that
the two members of a flux line pair are associated with plaquettes
of opposite circulation. Each of the models is invariant under the
transformation $p \rightarrow 1-p$, and so we consider only $0\leq
p \leq 1/2$. The CF and the O(1) models are both (statistically)
self-dual for all $p$ under a Kramers-Wannier transformation that
takes $\alpha$ to $\pi/2-\alpha$, leaving the line $\alpha=\pi/4$
invariant. The RBIM is not self-dual, except at $p=0$. Finally,
the CF and O(1) models are equivalent, under a gauge
transformation, on the line $p=1/2$.

Some of the differences in the behavior of these three models
can be illustrated by solving their one-dimensional versions,
which consist of a single chain of links and nodes.
In one dimension, disorder in the sign of the nearest-neighbor exchange
interaction can be removed from the RBIM by gauge transformation,
and the inverse localization length
has the disorder-independent value
$\xi_{\rm RBIM}^{-1}\equiv {\rm arctanh}(|\sin(\alpha)|)$,
finite for all $\alpha\not= 0,\pi/2$.
An elementary
calculation gives for the CF model
\begin{equation}
\xi_{\rm CF}^{-1} = |1-2p|\xi_{\rm RBIM}^{-1}\,,
\label{1d}
\end{equation}
so that $\xi_{\rm CF}$ diverges as $p \rightarrow 1/2$
but is otherwise finite, while for the O(1) model
$\xi^{-1}_{\rm O(1)}=0$ for all $p\not=0,1$.
Thus, the localization properties of the one-dimensional CF model at
$p\not= 1/2$ are like those of models belonging to
the Wigner-Dyson universality classes,
in that states are localized, while the absence of localization
in the O(1) model mirrors that found previously in quasi-one dimensional
class D systems \cite{bfgm}.

A second useful approach illustrating differences between these
models is to consider their two-dimensional versions in the limit
of weak inter-plaquette tunneling ($\alpha \ll 1$ or $\pi/2 -
\alpha \ll 1$) and weak disorder ($p\ll 1$). We do this in terms
of the discrete-time evolution operator $U$ in a closed system,
which is an $N_l \times N_l$ unitary matrix for a network of $N_l$
links \cite{km,hc}; a similar approach to a different free fermion
representation of the RBIM was used in Ref.\,\cite{bp}. The
eigenvalues of $U$ lie on the unit circle and may be written as
$e^{-i\epsilon}$. The real $\epsilon$ ($-\pi < \epsilon \leq \pi$)
play the role of excitation energy eigenvalues, and are
distributed symmetrically in pairs around $\epsilon=0$ because $U$
is a real orthogonal matrix. Long-time properties are determined
by the part of the spectrum near $\epsilon = 0$, on which we now focus.
At zero tunneling, it is sufficient to examine an isolated
plaquette. In our disorder-free reference system, the evolution
operator for a single plaquette satisfies $U^4=-1$, and hence
$\epsilon = \pm \pi/4, \pm 3\pi/4$. For a single plaquette with a
flux line added, $U^4=1$ and $\epsilon = 0,\pi, \pm \pi/2 $.
Turning on weak tunneling, it is clear that the spectrum near
$\epsilon = 0$ for a large system will arise by hybridisation of
the $\epsilon=0$ states from plaquettes with flux lines. In both
the RBIM and CF models, there are two scales for this
hybridisation, because flux lines appear in the system in adjacent
pairs associated with plaquettes of the same circulation. The
first consequence of tunneling is to remove the degeneracy within
each pair, yielding approximate eigenvalues $\epsilon=\pm
\epsilon_0$. At small $p$, pairs are dilute and tunneling between
different pairs is  not sufficient to generate extended states at
$\epsilon =0$. By contrast, for the O(1) model in this regime,
there is only one scale for hybridisation, since single flux lines
appear independently on the set of weakly-coupled plaquettes. As a
result, metallic behavior is not excluded even at $p,\alpha \ll1$.

Our results from numerical simulation supplement this qualitative
discussion. We study the CF and O(1) models in cylindrical
geometry via the transfer matrix $T$, obtaining the positive
Lyapunov exponents, $0\leq \nu_1 \leq \dots \leq \nu_{M}$ in a
system of width $M'=2M$ links. A crucial technical aspect of these
calculations is our discovery that the standard algorithm
\cite{mk,ps} has a serious instability to roundoff errors
throughout much of the phase diagram of both models. More
specifically, we find that the smallest positive Lyapunov
exponent, $\nu_1$, may be either identically zero or exceptionally
small ($\nu_1 \ll M^{-1}$). (The first happens in the O(1) model
for all $p$ and $\alpha$, and in the CF model on the self-dual
line $\alpha=\pi/4$; the second happens in the metallic phase of
the CF model.) Under these circumstances, numerical noise from
roundoff errors generates a systematic positive error in the value
obtain for $\nu_1$. From an analytical theory \cite{c} of the
instability, we find that the error in $\nu_1$ decreases with
reduced noise amplitude $\eta$ only as $|\log(\eta)|^{-1}$. This
instability can be cured by making explicit use in numerical
calculations of the structure imposed on $T$ by current
conservation and the symmetry of class D.

In detail, $T$ has the polar decomposition
\begin{equation}
T = \left( \begin{array}{cc}
A_1 & 0 \\
0 & A_2
\end{array} \right)
\left(\begin{array}{cc}\cosh(\gamma) & \sinh(\gamma)\\
\sinh(\gamma) & \cosh(\gamma) \end{array} \right)
\left(\begin{array}{cc}
A_3^{T} & 0\\ 0 & A_4^{T} \end{array} \right) \,,
\label{T}
\end{equation}
where $A_1 \ldots A_4$ are $M \times M$ real orthogonal matrices
and $\gamma$ is an $M\times M$ real diagonal matrix. It follows
that $T^{T}T$ is diagonalized by the transformation
$B^{T}T^{T}TB$, where
\begin{equation}
B=\left(\begin{array}{cc}
A_3 & A_3\\
A_4 & -A_4 \end{array} \right) \,.
\label{B}
\end{equation}
The standard method for calculating Lyapunov exponents numerically
involves acting with the transfer matrices for successive slices
of the system on a set of $M$ orthogonal vectors, and reimposing
orthogonality by means of Gram-Schmidt transformations
\cite{mk,ps}. If all Lyapunov exponents are separated by gaps,
this set of vectors converges to the eigenvectors of $T^{T}T$
associated with the first $M$ exponents. Convergence rates are
determined by the sizes of gaps between successive exponents. In
the present case, convergence rates are seriously reduced if
$\nu_1$ approaches zero, so that the gap between the smallest
positive and largest negative exponents vanishes. Moreover,
numerical noise ultimately limits the extent of convergence,
and leads to an erroneously large value for $\nu_1$.
To overcome this, we impose on the $M$ vectors concerned not simply
orthogonality but instead the fact that their first $M$ components
separately form an orthogonal matrix $A_3$, and their last $M$
components form $A_4$, as is evident from Eq.\,\ref{B}. The
results we obtain in this way for the CF model differ
significantly from those of Ref.\,\cite{cf}.

\begin{figure}
\epsfxsize=3.5in \centerline{\epsffile{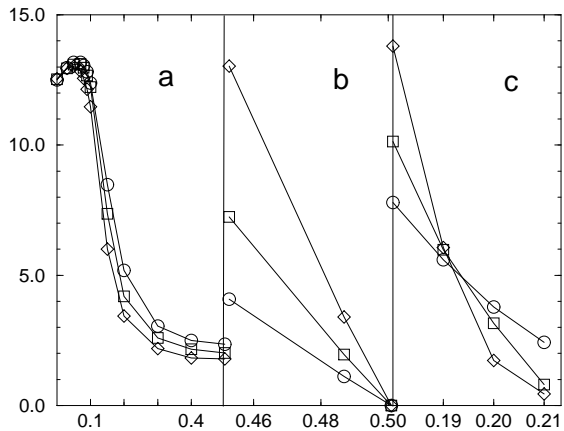}}
\vspace{0.0in} \caption{Behavior of the CF model in systems of
width $M'=64$ ($\circ$), $128$ ($\Box$), and $256$ ($\Diamond$).
(a) The self-dual line: $M'\nu_2$ as a function of $p$. (b)
Quantum-Hall--type transition: $M'\nu_1$ as a function of
$\sin^2(\alpha)$ at $p=0.1$. (c) Insulator -- metal transition:
$M'\nu_1$ as a function of $p$ at $\sin^2(\alpha)=0.19$.}
\label{combined}
\end{figure}

Evidence in support of the phase diagram of Fig.\,\ref{phasediag}
for the CF model is presented in
Fig.\,\ref{combined}. On the self-dual line ($\alpha=\pi/4$) we
believe that $\nu_1$ is identically zero (as in the
one-dimensional O(1) model). For example, at $p=1/2$ and
$\alpha=\pi/4$ we obtain in systems of length $L=5\cdot 10^5$ the
bounds $\nu_1 < 1.5\cdot 10^{-3}$ at width $M'=4$ and $\nu_1 < 1.5
\cdot 10^{-4}$ at $M'=256$. In order to search for a possible
multicritical point on the self-dual line, we therefore examine
the behavior of $\nu_2$ \cite{footnote}. If there is a
multicritical point at $p=p_{\rm MC}$, one expects the amplitude
ratio $M'\nu_2$ to show three regimes at large $M'$, as a function
of $p$. For $p < p_{\rm MC}$, scaling flow is towards smaller $p$
and $M'\nu_2$ has a $p$-independent value governed by the critical
point at $p=0$. At $p=p_{\rm MC}$, a distinct limiting value
arises from the multicritical point. And for $p > p_{\rm MC}$,
scaling flow of the conductivity in the metallic phase towards
larger values means that $M'\nu_2$ will slowly decrease
towards zero with increasing $M'$. The data shown in
Fig.\,\ref{combined}a are consistent with this scenario, although
the position of the multicritical point is not well-determined: we
find the bounds $0.05 \leq p_{\rm MC} \leq 0.15$. A
quantum-Hall--type transition is observed on crossing the
self-dual line by varying $\alpha$ at fixed $p<p_{\rm MC}$, as
illustrated in
Fig.\,\ref{combined}b: $M'\nu_1$ increases with $M'$ for $\alpha
\not= \pi/4$ (localization) and vanishes as $\alpha \rightarrow
\pi/4$ (delocalization). (This transition is expected \cite{sf2}
to be in the universality class of the pure Ising transition,
because the disorder strength scales towards zero, as in the RBIM
at small $p$.) We determine the position of the metal-insulator
phase boundary from the variation of $M'\nu_1$ with $p$ and $M'$
at fixed $\alpha\not= \pi/4$, shown in
Fig.\,\ref{combined}c. Since $M'\nu_1$ decreases rapidly with
increasing $M'$ in the metal and increases with $M'$ in the
insulator, the critical point, $p_{\rm C}(\alpha)$, is identified
by the crossing of curves for different $M'$. In this way, we
arrive at the phase diagram for the CF model displayed in
Fig.\,\ref{phasediag}.

We believe that the O(1) model has only a metallic phase, and has
$\nu_1$ identically zero for all $p\not= 0$. Our calculations
cover the range $0.1\leq p < 0.5$ and $0.1 \leq \sin^2(\alpha)
\leq 0.5$. If the model were to support a localized phase, it
should appear at small $p,\alpha$. As an illustration of the
absence of localization, at $p=0.1$, $\sin^2(\alpha)=0.1$, we
calculate for $M'=16$: $\nu_1 < 10^{-3}$ in the O(1) model, while
$\nu_1 = 0.83$ in the CF model.

In summary, we find that two-dimensional models for localization
in the symmetry class D can have quite different behavior
according to the form of disorder. Several additional points
deserve emphasis. The metallic phase of the CF model is self-dual,
as is its multicritical point. By contrast, the RBIM is not
self-dual but has higher supersymmetry at its multicritical point
\cite{grl}. There is little reason to suppose that these two
multicritical points are in the same universality class.
Separately, the apparent absence of an insulating phase in the
O(1) model is remarkable, because the bare conductivity becomes
small when $\alpha \to 0$ or $\pi/2$. Recently, it has been
emphasized that the target manifold of the class D nonlinear sigma
model is not connected \cite{bsz}, and this means that domain wall
excitations can occur in the sigma model, which must be described
by additional parameters, and have not been taken into account in
weak-coupling analyses so far. It is likely that these domain
walls in the sigma model language are connected with the richness
of phases in this symmetry class. In that context, the O(1)
model with $p=1/2$ is known to be a special case, since it
maps to a sigma model without domain walls: this
fact suggests that proliferation of domain walls may be necessary for
localization \cite{bsz,grl,rl}.

The work was supported in part by: the EPSRC under 
Grant GR/J78327 (JTC); the NSF under Grants DMR-98-18259
(NR) and DMR-00-75064 (AWWL); and the DIP German Israeli program (BH and YA).


\vspace*{-5mm}


\begin{references}


\vspace*{-15mm}



\bibitem{az} A. Altland and M.R. Zirnbauer, Phys. Rev. B {\bf 55}, 1142
(1997); M.R. Zirnbauer, J. Math. Phys. {\bf 37}, 4986 (1996).


\bibitem{sfbn} T. Senthil {\it et al.}, Phys. Rev. Lett. {\bf 81}, 4704 (1998).



\bibitem{khac} V. Kagalovsky {\it et al.}, \prl {\bf 82}, 3516
(1999).

\bibitem{brad} T. Senthil, J.B. Marston, and M.P.A. Fisher, \prb {\bf 60},
4245 (1999);
I.A. Gruzberg, A.W.W. Ludwig, and N. Read, \prl {\bf 82}, 4254
(1999).

\bibitem{bfgm} P.W. Brouwer {\it et al.}, \prl {\bf 85}, 1064 (2000).

\bibitem{bundschuh}
R. Bundschuh {\it et al.}, \prb {\bf 59}, 4382 (1999).

\bibitem{sf2}
T. Senthil and M.P.A. Fisher, \prb {\bf 61}, 9690 (2000).

\bibitem{readgr} N. Read and D. Green, \prb {\bf 61}, 10267 (2000).

\bibitem{bsz} M. Bocquet, D. Serban, and M.R. Zirnbauer,
Nucl. Phys. B {\bf 578}, 628 (2000).

\bibitem{grl} I.A. Gruzberg, N. Read, and A.W.W. Ludwig, cond-mat/0007254.

\bibitem{cc} J.T. Chalker and P.D. Coddington, J. Phys. C {\bf 21}, 2665
(1988).

\bibitem{cf} S. Cho and M.P.A. Fisher, Phys. Rev. B {\bf 55}, 1025
(1997).

\bibitem{chothes} S. Cho, PhD thesis, UC Santa Barbara, 1997,
unpublished.

\bibitem{rl} N. Read and A.W.W. Ludwig, cond-mat/0007255.

\bibitem{km} R. Klesse and M. Metzler, Europhys. Lett.{\bf 32}, 229(1995).

\bibitem{hc} C.-M. Ho and J.T. Chalker, Phys. Rev. B {\bf 54} 8708 (1996).

\bibitem{bp} J.A. Blackman and J. Poulter, Phys. Rev. B{\bf 44}, 4374
(1991).

\bibitem{c} J.T. Chalker {\it et. al.}, unpublished.

\bibitem{mk} A. MacKinnon and B. Kramer, Phys. Rev. Lett. {\bf 47},
1546 (1981); Z. Phys. B{\bf 53}, 1 (1983).

\bibitem{ps}J.L. Pichard and G. Sarma, J. Phys C {\bf 17}, 4111 (1981).

\bibitem{footnote} A complication arises from the fact that the CF
model is not (statistically) invariant under $90^{\circ}$
rotations, except at $p=0$ and $p=1/2$. Because of this ($p$ and
$\alpha$ dependent) anisotropy, the $\nu_n$ have been calculated
throughout as the geometric mean of the Lyapunov exponents for
cylinders with perpendicular orientations relative to the lattice.



\end{references}
\end{document}